\documentclass[prl,twocolumn,showpacs,superscriptaddress,preprintnumbers,amssymb,hyperlink]{revtex4-1}
\usepackage{tabularx,xcolor}
\usepackage{graphicx,adjustbox} 
\usepackage{amsmath}
\usepackage{soul}

\newcommand{\beq}{\begin{equation}}
\newcommand{\eeq}{\end{equation}}
\newcommand{\beqn}{\begin{eqnarray}}
\newcommand{\eeqn}{\end{eqnarray}}

\begin{document}

\title{Density wave halo around anyons in fractional quantum anomalous hall states}
\author{Xue-Yang Song}
\author{T. Senthil}
\affiliation{Department of Physics, Massachusetts Institute of Technology, Cambridge, Massachusetts 02139, USA}
\date{\today}

\begin{abstract}
    The recent observation of fractional quantum anomalous Hall (FQAH) states in tunable moir{\'e} materials encourages study of several new phenomena that may be uniquely accessible in these platforms.  Here, we show that an isolated localized anyon of the FQAH state will nucleate a `halo' of Charge Density Wave (CDW) order around it. We demonstrate this effect using a a recently proposed quantum Ginzburg-Landau theory that describes the interplay between the topological order of the FQAH and the broken symmetry order of a CDW. The spatial extent of the CDW order will, in general, be larger than the length scale at which the fractional charge of the anyon is localized. The strength and decay length of the CDW order around anyons induced by doping or magnetic field differs qualitatively from that nucleated by a random potential. Our results leverage a precise mathematical analogy to earlier studies of the superfluid-CDW competition of a system of lattice bosons which has been used to interpret the observed CDW halos around vortices in high-$T_c$ superconductors. We show that measurement of these patches of CDW order can give an indirect route to measuring the fractional charge of the anyon.  Such a measurement may be possible by scanning tunneling microscopy (STM) in moir{\'e}  systems. 
\end{abstract}

\maketitle

\section{Introduction}
The (fractional) quantum Hall effect usually occurs in the setting of two dimensional electron gases subject to a strong magnetic field. It has been clear for some time now that quantum Hall physics can be realized without a magnetic field if electrons either completely or partially fill\cite{sun2011nearly,sheng2011fractional,neupert2011fractional,wang2011fractional,tang2011high,regnault2011fractional,bergholtz2013topological,parameswaran2013fractional} a Chern band in a lattice system. When the required time reversal symmetry breaking occurs spontaneously, the resulting phases  have been dubbed (Fractional) Quantum Anomalous Hall states. In the last few months, Fractional Quantum Anomalous Hall (FQAH)  states have been discovered in two different two dimensional moire heterostructures\cite{cai2023signatures,zeng2023integer,park2023observation,lu2023fractional}, following directions suggested by a number of theoretical studies\cite{zhang2019nearly,ledwith2020fractional,repellin2020chern,abouelkomsan2020particle,wilhelm2021interplay,wu2019topological,yu2020giant,devakul2021magic,li2021spontaneous,crepel2023fci}. 

These FQAH states compete with a number of other states of matter. At the lattice fillings $p/(2p+1)$ (for $p$ an integer), where current experiments see an FQAH state, a natural competing insulating state is a commensurate Charge Density Wave (CDW) insulator. Indeed it has been possible to tune out of the FQAH phase into insulating phases with large longitudinal (and small Hall) resistivity using a perpendicular displacement field. The competition/intertwinement between the FQAH and CDW potentially manifests itself in ways that are unique to the lattice setting, and are absent in the continuum realization of quantum hall physics studied experimentally over the previous decades. 

\begin{figure}
\adjustbox{trim={.18\width} {.15\height} {.1\width} {.1\height},clip}
   {\includegraphics[width=.8\textwidth]{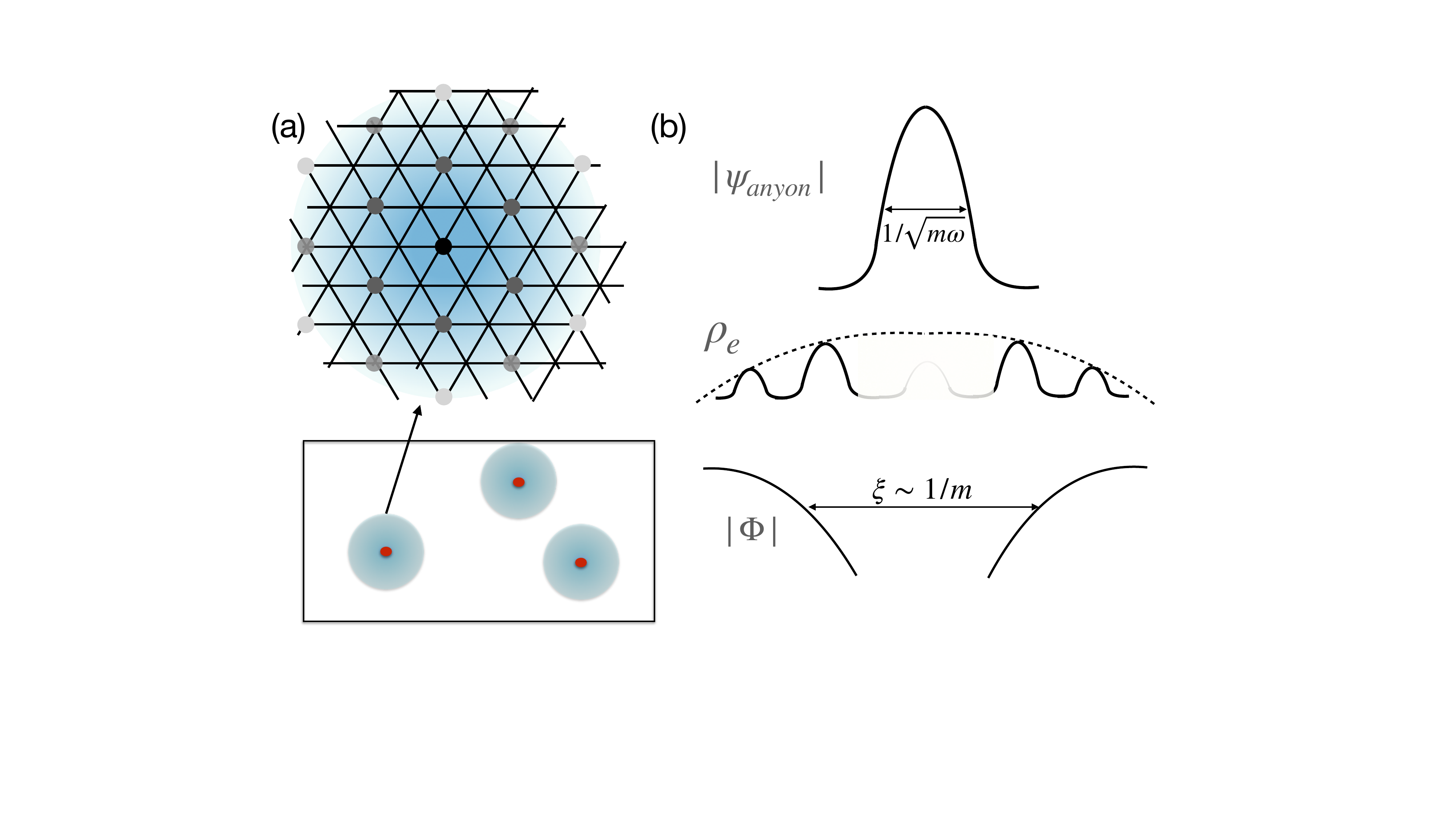}}
    \caption{(a): The charge density pattern around an anyon `core'. The gray scale is correlated with the intensity of the CDW order which decays exponentially in space. Lower panel shows the FQAH phase with a finite density of anyons (big disks), where the anyon fractional charge (red small disks) is more localized than the surrounding CDW halo. (b):(Top to bottom) Schematic plot of the anyon charge density, the charge density wave,  and the composite boson density. The spatial region where the fractional charge of the anyon is localized is related to the quantum zero point motion of the vortex (of the composite boson superfluid) while the induced CDW order decays at a `coherence' scale $xi$ related to the distance to the critical point.  The composite boson density is depleted at the same coherence length scale. }
    \label{Fig:CDW}
\end{figure}

Recent theoretical work\cite{song2023phase}, which we briefly review below,  has derived a quantum Ginzburg-Landau (q-GL) theory to describe the competition/intertwinement between the FQAH and CDW state, and used it to address possible phase transitions (and phases proximate to the FQAH state). Much like in other instances of the use of  such theories, we can use the q-GL theory to address many other questions pertinent to the interplay between FQAH and CDW physics. 

In this paper we show that in the FQAH phase, a localized anyon nucleates around it a halo where there is CDW ordering. The size of the halo region   is determined by the ``distance" to the fully ordered CDW phase in the phase diagram (assuming the FQAH-CDW transition is at least approximately second order). In a device where a finite (but low) density of anyons are present, perhaps due to a small doping away from the filling $p/(2p+1)$, there will be a corresponding finite density of CDW patches. If the inter-anyon separation is larger than the halo size (which is guaranteed at sufficiently small doping), then the CDW patches will be well separated from each other. A schematic depiction is in Fig. \ref{Fig:CDW}.

Such CDW patches may in principle be imaged through future STM experiments. We point out that such an imaging can be an indirect measure of the anyon charge. For instance at filling $\nu = \frac{p}{q} + \delta$, suppose the anyons that are induced by the doping have a charge $q*$ (for odd denominator states, $q^*$ will be a multiple of $\frac{1}{q}$). Then there will be $\delta/q^*$ CDW patches per unit cell. Thus the ratio of CDW patch density to the doping density within the FQAH plateau directly gives the fractional charge $q^*$ of the anyon. 

The phenomenon of CDW order nucleating in a halo around the anyon is analogous to the observation of CDW order in a halo around a vortex core in cuprate superconductors\cite{hoffman_STM}. In the latter system too, this arises due to a competition between superconducting and CDW orderings.  This competition was discussed theoretically in simplified models of bosonic Cooper pairs at fractional lattice density in a series of papers\cite{lannert2001quantum,balents2005putting,balents_competing}. The resulting q-LG theories for the superfluid-CDW Mott transition of bosons at fractional lattice filling enable addressing the physics of the vortex structure in the superfluid state and the nucleation of CDW order once the superfluid order parameter is suppressed near the vortex core\cite{balents2005putting,balents_competing}. 

The q-LG theory for the FQAH-CDW competition in our previous work is a modification of the q-LG theory for the superfluid-CDW competition of bosons on a lattice. To explain this connection, we recall that, in the usual Landau level setting,  there is a Ginzburg-Landau description\cite{zhang1989effective,read1989order} of the fractional quantum Hall system in terms of ``composite bosons" (electrons with an odd number of attached vortices) coupled to emergent gauge fields with a Chern-Simons action. In this description, the quantum Hall phase is a condensate (a `superfluid') of composite bosons. The theory of Ref. \cite{song2023phase}  described the FQAH state at any lattice filling  in a generalization of this standard  composite boson picture. The composite bosons are at the same lattice filling as the electron. Condensing the composite boson gives the FQAH phase while a CDW Mott insulator of these composite bosons is a possible proximate electronic CDW insulator. Thus there is a close relationship between the theory of the FQAH-CDW competition and that of the superfluid-CDW competition, though the physical interpretation is different. 

Interestingly for our purposes, the vortices of the composite boson condensate are precisely the fractionally charged anyons of the FQAH state. Thus the earlier discussion\cite{balents_competing} of the nucleation of CDW around vortices in the superfluid phase of bosons on a lattice can be essentially taken over directly to reach the conclusions outlined above on CDW halos around anyons in the FQAH state. We also study defects in the CDW ordered phase, and show that, apart from having fractional charge, there are circulating electrical currents in their vicinity that reflects the proximate FQAH phase. 

In the rest of this paper we will flesh out the details of these introductory remarks.

\section{Quantum Ginzburg-Landau theory for FQAH-CDW competition}
To be concrete, we specialize to lattice filling $\nu = -\frac{2}{3}$ where the most prominent FQAH state is seen in t-MoTe$_2$. It is convenient to go to a hole picture and describe this FQAH state as having the same topological order as the Laughlin state of holes at $1/3$ filling.  The phase transition out of the quantum Hall phase can be thought of as a superfluid-insulator transition of  (generalized) composite bosons $\Phi$ at a lattice filling of $1/3$. The theory is conveniently described\cite{song2023phase} in terms of vortices $\Phi_{vI}$ (with $I = 1, 2,3$) of the composite boson field,  with a  Lagrangian\cite{song2023phase}   
\begin{equation} 
\label{1/3fqhtocdw}
 {\cal L}  =  \sum_{I = 1}^3 |\left(\partial_\mu - i \alpha _\mu \right)\Phi_{vI}|^2 +{\mathcal L}_4 +\cdots  - \frac{3}{4\pi} \alpha d\alpha + \frac{1}{2\pi} Ad\alpha,
\end{equation}
The ellipses represent terms consistent with the translation and $C_3$ symmetries of the lattice. The $\alpha$ is a dynamical $U(1)$ gauge field and $A$ is a probe $U(1)$ gauge field that can be used to calculate the electromagnetic response. We omitted an additional term  $1/(4\pi)AdA$ describing an integer quantum anomalous Hall effect of the filled Chern band, which does not affect the main discussion. 

The symmetry action of translations and $C_3$ (the subscript $I+1$ are defined modulo $3$, i.e. $I \mod 3 +1$)\cite{burkov2005superfluid} reads,
\begin{eqnarray}
\label{eq:vortexsym}
    T_1&:& \Phi_{v,I}\rightarrow \Phi_{v,I+1}\nonumber\\
    T_2&:&\Phi_{v,I}\rightarrow e^{i\frac{2(I+1)\pi}{3}} \Phi_{v,I+1}\nonumber\\
    C_3&:&\Phi_{v,1}\rightarrow e^{i\frac{\pi}{6}} \Phi_{v,3},\Phi_{v,2}\rightarrow -i\Phi_{v,1},\nonumber\\
    &&\Phi_{v,3}\rightarrow e^{i\frac{\pi}{6}}\Phi_{v,2}.
    \end{eqnarray}
    The allowed quartic terms for the dual vortex fields have the structure
\begin{eqnarray}
    \mathcal L_4[\Phi_{vI}]& = & {\cal L}_u + {\cal L}_v \nonumber \\
    {\cal L}_u & = & u(\sum_{I=1}^3 |\Phi_{vI}|^2)^2 \nonumber\\ 
    {\cal L}_v & = & v(|\Phi_{v1}|^2|\Phi_{v2}|^2+|\Phi_{v1}|^2|\Phi_{v3}|^2+|\Phi_{v2}|^2|\Phi_{v3}|^2), \nonumber
\end{eqnarray}
where the sign of $v$ determines the nature of the symmetry breaking state when $\Phi_{vI}$'s condense.

Clearly if the $\Phi_{vI}$ are all gapped, we get the Laughlin $1/3$ state. If however the $\Phi_v$ condense, then the $\alpha$-field is Higgsed and the quantum hall effect is destroyed. The low energy long wavelength response to the probe $A$ field is then given by a Maxwell action (with no Chern-Simons term). Thus we get a CDW insulating state, where the non-trivial translation action on $\Phi_{vI}$ ensures that translation symmetry is spontaneously broken. The pattern of charge ordering in the CDW is determined by the structure of the non-linear terms  in Eq. \eqref{1/3fqhtocdw}\cite{burkov2005superfluid}.

Let $\mathbf{R}_{1,2}$ be the primitive vectors of the real space moir\'e superlattice, and $\mathbf{b}_{1,2}$ be the corresponding reciprocal lattice vectors. The charge density operator at a wavevector $\frac{2m}{3}\mathbf{b}_1+\frac{2n}{3} \mathbf{b}_2$ can  be written in terms of the low energy vortex fields as (where again the index $j$ on vortex field is to be interpreted as $(j-1) \textrm{mod }3+1$)
\begin{eqnarray}
\label{eq:rhomn}
\rho_{mn}(\mathbf{R} ) & = & S(m,n) \sum_{j=1,2,3} \Phi_{v,j}^\dagger(\mathbf{r}) \Phi_{v,j+n}(\mathbf{r}) e^{i(n-m)j\frac{2\pi}{3}},\nonumber\\
\rho(\mathbf{R})& = & \sum_{m,n}\rho_{mn}(\mathbf{R}) e^{i\frac{2\pi m}{3}\mathbf{R}_1+i\frac{2\pi n}{3}\mathbf{R}_2},
\end{eqnarray}
where $S(m,n)$ is a form factor unknown from the low energy properties of the vortex fields.
It is straightforward to verify that the vortex transformation in Eq \eqref{eq:vortexsym} determines that the $\rho_{mn}$ sits at the correct wavevector.

Thus the theory in Eq.~\eqref{1/3fqhtocdw} enables discussion of the competition between the FQAH and CDW phases within a continuum field theory. It can thus be thought of as a quantum Ginzburg-Landau theory that is suitable for discussing universal aspects of the physics. We can use it to study  many different aspects of the interplay between the two kinds of orders, such as, eg, their phase transition, and inhomogenous situations involving interfaces between the FQAH and the CDW. Here however we will study the structure of the charged anyon defects of the FQAH phase.   

For simplicity, we assume that the charge-density-wave order is obtained by condensing only one of the $\Phi_v$ fields, which happens when $v>0$. The resulting CDW order pattern is the simplest one for $1/3$ filling on triangular lattices -- with a tripled unit cell shown in Fig. \ref{Fig:CDW}. Other condensation patterns in Ref. \onlinecite{burkov2005superfluid} are allowed for different signs of $v,u$ in $\mathcal L_4$, though this does not affect the discussion below, with the only difference being the detailed CDW pattern nucleated by an anyon.

\section{Intertwinement of FQAH and CDW phases}

The FQAH state is formed by `condensing' the composite bosons $\Phi$ which is equivalent to considering a phase where the vortex fields $\Phi_{vI}$ are gapped. A vortex defect, localized at some spatial point, in the composite boson condensate is thus a unit charged source for the $\alpha$ gauge field in Eq.  \eqref{1/3fqhtocdw}.     The Chern-Simons term binds a magnetic flux of $2\pi/3$  of $\alpha$ to the localized $\Phi_{vI}$ defect.  The coupling to the probe gauge field $A$ then tells us that this defect has charge-$1/3$ and is  identified with the $1/3$ Laughlin quasiparticles. 

The intertwinement of the two phases is manifested by the defect of `order parameters' in one phase carrying quantum numbers of the order parameters of the other phase. In the FQAH phase, the defect of composite boson superfluid, i.e. the vortices $\Phi_v$, transforms in a projective representation of the space group, which directly leads to the charge density wave pattern, obtained  when the vortices  condense.   In particular, $\Phi_v$ obeys $T_1T_2T_1^{-1}T_2^{-1}=e^{i2\pi/3}$ (the magnetic translation due to $1/3$ filling of the bosons). The CDW pattern will have a minimal $3$-site unit cell to trivialize the magnetic translation of the vortices. 
Fig. \ref{Fig:CDW} shows a schematic CDW nucleated around a vortex. The amplitude of the CDW, denoted by the gray scale of the disks, decays exponentially away from the vortex core. Hence we arrive at a remarkable conclusion: a localized $e/3$ anyon in the FQAH phase will nucleate CDW orders at its `core', indicative of the symmetry breaking competing phase  proximate to the FQAH.

\begin{figure}
\adjustbox{trim={.24\width} {.36\height} {.2\width} {.16\height},clip}
   {\includegraphics[width=1.\textwidth]{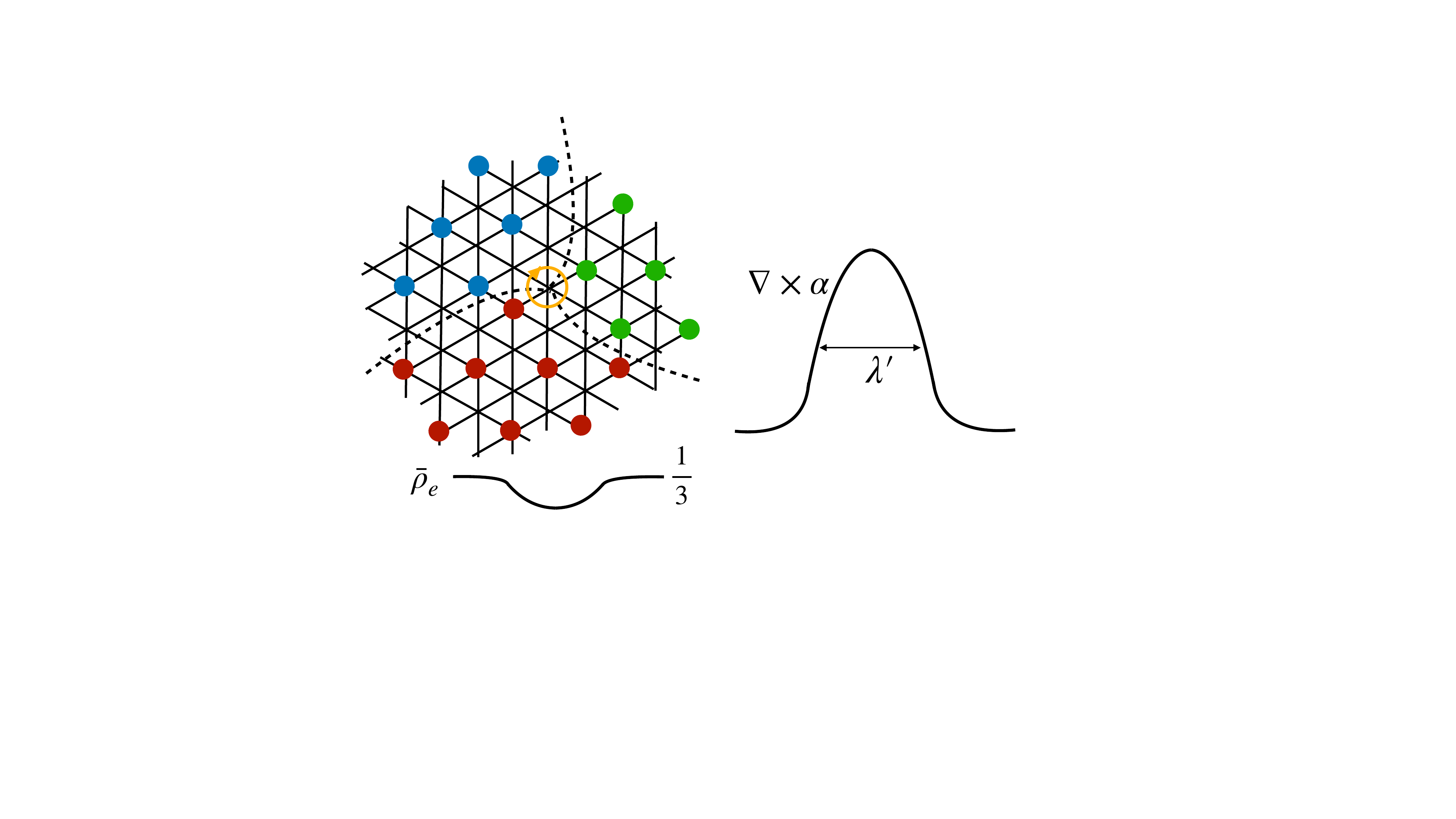}}
    \caption{Left: The defect in the CDW ordered phase where a $-1/3$ charge deficit as drawn of an empty site. The dashed lines are the domain walls of the CDW order parameter: the orders on two sides of the wall are related by translation by one lattice constant along the domain wall direction. The charge patterns in different domains are distinguished by colors.   Around the defect, the coarse-grained  charge density $\bar\rho_e$ will deviate from the average $1/3$ in the CDW phase.  A circulating current around the defect, i.e. an FQAH patch, will change the local magnetization. Right: The defect is identified as a flux of the gauge field $\alpha$, with a spatial  extent given by the penetration depth of the vortex condensate. }
    \label{Fig:defect}
\end{figure}

On the other hand, in the CDW phase when $\Phi_v$ condenses, we can see traces of the FQAH in the defect of the CDW order. In Fig \ref{Fig:defect} we schematically show a defect of the CDW order, which carries $-1/3$ or $2/3$ excess charge upon careful counting. From the field theory for the closely related superfluid-insulator transition at this lattice filling, it has been argued that the bosons will be fractionalized near the critical point into $3$ fundamental fields each carrying $1/3$ of the boson charge\cite{balents_competing}. This is because a $2\pi$ dual `vortex' of any $\Phi_v$ field induces $2\pi/3$ of $d\alpha$ flux, and hence carries $1/3$ of boson charge.  The same argument continues to hold in our theory with the extra Chern-Simons term for $\alpha$. As we show below,  the Chern-Simons term leads to the presence of circulating electrical currents which will lead to a change of the local magnetization. 

 We identify the defect as a  vortex in the $\Phi_v$ superfluid.  An effective  Lagrangian for the superfluid as
\begin{align}
    \mathcal L_{CDW}=\frac{\rho_s}{2}|(\nabla \theta-\alpha)|^2+\frac{3}{4\pi}\alpha d\alpha\nonumber\\
    +\frac{g^2}{8\pi}(-b^2+e^2),
\end{align}
where $\Phi_v \sim e^{i\theta}$, and $\rho_s$ is the corresponding  ``superfluid density". The last term is the Maxwell term for the internal gauge field $\alpha$ with $e=-\nabla \alpha_0$ (we fix the gauge to be \emph{time independent}). $g^2$ is a coupling constant with the dimension of length. The equations of motion
\begin{eqnarray}
    -g^2\nabla\times b&=&-4\pi\rho_s(\nabla \theta-\alpha)+6 \vec e\times \hat z,\nonumber\\
    \frac{3b}{2\pi}&=&\frac{g^2\nabla \cdot e}{4\pi},
\end{eqnarray}
correspond to modified Maxwell equations due to the Chern Simons term, where $\left(\vec e\times \hat z\right)_i=\epsilon_{ij}e_j$. We consider a $2\pi$ vortex of $\Phi_v$ located at the origin. Then $\nabla \times \nabla \theta = 2\pi \delta^2(r)$. We take a curl for the first equation,  and substitute $\vec e=-\nabla \alpha_0$ to arrive at the modified vortex equation
\begin{eqnarray}
    \nabla^2(b-\frac{6\alpha_0}{g^2})-\frac{b}{\lambda^2}&=&\frac{1}{2\lambda^2}\delta^2(r),\nonumber\\
    6b&=&g^2\nabla^2\alpha_0,
\end{eqnarray}
where the penetration length $\lambda=|g|/\sqrt{4\pi\rho_s}$. Using the second equation and eliminating $\alpha_0$ in the first equation, we recover the usual London equation for a vortex in a BCS superconductor, with a modified penetration length 
\begin{align}
    \frac{1}{\lambda'^2}=\frac{1}{\lambda^2}+\frac{36}{g^4}.
\end{align}

As expected, the vortex in the $\Phi_v$ condensate traps a flux of $b$ localized to a length scale $\lambda'$ around it.   From the modified Gauss law $\frac{3b}{2\pi}=\frac{g^2\nabla \cdot e}{4\pi}$, it follows that a defect induces a radial electric field $e$. Since $d\alpha$ is coupled to the physical probe field $A$, the physical current is given by $\vec J=\vec e\times \hat z$. So a circulating current appears around the defect of the CDW phase.

We see that the role of the Chern-Simons term is to lead to a circulating electrical current near the defect core which will lead to a change of the orbital magnetization compared to the bulk regions far away from the defect. A heuristic picture is that in the vicinity of the defect the CDW order is suppressed and we instead nucleate a small bubble of the FQAH phase. Thus there is an interface created, at a scale of the coherence length, between FQAH and CDW which then leads to a circulating electric current. We thus see that the $e/3$ Laughlin quasiparticles imprint on the defect of the CDW order parameters $|\Phi_v|$ in the CDW phase.

\section{Density wave halo around anyons}

We now study in more detail the spatial structure of  an anyon defect in the FQAH phase and the halo of charge density modulation. This charge  modulation pattern is indicative of the CDW phase on the other side of the transition, obtained e.g. by tuning the displacement field in the dual-gate moire TMD system. This phenomenon is reminiscent of  the modulation of the local density of states found\cite{hoffman_STM} near the vortex cores in superconducting high $T_c$ cuprates, which  has been  attributed to the intertwinement with charge ordering in the underdoped regime originating from proximity to the Mott insulator.

For our purposes it is sufficient to work with an approximate version of Eq. \eqref{1/3fqhtocdw} where we retain only quadratic terms in  $\Phi_{vI}$. This will be legitimate so long as we are close enough to the phase transition to warrant using a long wavelength theory but not so close as to require a full-blown solution of the interacting field theory. 
In this regime, the effective action for the vortex  can be taken to have  the form
\begin{align}
\label{eq:vortexaction}
    S=\int d^2 r d\tau \sum_{I = 1}^3 |\left(\partial_\mu - i \alpha _\mu \right)\Phi_{vI}|^2 + m^2|\Phi_{vI}|^2+\mathcal L_4\nonumber\\
    - \frac{3}{4\pi} \alpha d\alpha + \frac{1}{2\pi} Ad\alpha,
\end{align}
with the transition tuned by the mass $m$ (here $m$ is real in the FQAH). We have renormalized the spatial and temporal scales to set the velocity $v=1$. To leading order, we ignore the fluctuation of $\alpha_i$  . $\alpha_0$, however is needed to account for the vortex chemical potential, i.e. magnetic field effects and mediate vortex interactions.

We discuss three ways to induce the  anyon defect in the system: first through an impurity potential, second through doping within the quantum Hall plateau, and third by turning on a small magnetic field. 

\textbf{Impurity pinning:}
A static impurity potential couples linearly to the charge density through the form 
\begin{align}
    H_i= \int d^2 r\sum_{mn} V_{mn}(r)\rho_{mn}(r).
\end{align}
Linear response theory gives an induced charge density wave in terms of the impurity potential as\cite{balents_competing}
\begin{align}
    \langle \rho_{mn}(r)\rangle=S(m,n)^2 \int d^2r' \Pi_0(r-r')V_{mn}(r'),\nonumber\\
    \Pi_0(r)\approx -\frac{3m}{4\pi^2}\frac{e^{-2mr}}{(2mr)^{3/2}},\textrm{ as }r\rightarrow \infty,
    \label{eq:cdw}
\end{align}
Hence an impurity localized at some spatial position $\mathbf{r}_0$ induces a  charge density wave that decays exponentially from $\mathbf{r}_0$  over a characteristic length $\xi\sim 1/(2m)$ related to the distance to the critical point. This length scale can either be viewed as the correlation length for CDW fluctuations, or, equivalently, as  the coherence length of the superfluid of the composite bosons.   Fig \ref{Fig:CDW}(a) schematically shows the CDW modulation with a tripled unit cell induced by an impurity (blue disk). The detailed modulation pattern depends on microscopic details of the impurity and energetics etc.

As an impurity weakens the composite boson superfluidity, it is likely for the impurity site to overlap with a vortex core.  An important difference from the analogous discussion of Ref. \onlinecite{balents_competing}  is that the vortex  carries $e/3$ physical electrical charge. Hence the impurity potential will couple directly to the vortex density as a chemical potential $-|\mu|$. When the impurity strength is strong enough to overcome the vortex energy gap ($|\mu|>m$), an anyon is nucleated at the impurity location.  In the case considered  in  Ref \onlinecite{balents_competing}, as the vortex does not carry charge,  the impurity potential mainly only couples to the CDW order $\rho_{mn}$ with nonzero momenta. In contrast  here, the impurity can also act directly as a chemical potential for the vortex, which is similar to the case of a nonzero magnetic field in Ref. \onlinecite{balents_competing}.

To investigate the spatial profile of the pinned anyon, we approximate the impurity potential close to its minimum in space by a harmonic expansion:
\begin{align}
\label{eq:harmonic}
    i\alpha_0(\vec r)=-\mu+\frac{1}{2}m\omega^2r^2 \quad(\mu>0).
\end{align}

 We work in the limit $|i\alpha_0-m|\ll m$ and  ignore the vacuum fluctuations of vortex-antivortex pairs . We can then perform an expansion  to the lowest order of $\tau$- derivative  and $|i\alpha_0-m|$.  Following the procedure of  Ref \onlinecite{balents_competing}, define $\Phi_{v,l}=\frac{1}{\sqrt{2i\alpha_0}}\Psi_l$ and the low energy action reads
\begin{align}
\label{eq:harmonicaction}
    S=\int d^2r d\tau \sum_{I=1}^3 [\Psi^*_l\frac{\partial \Psi_l}{\partial\tau}+\frac{|\partial \Psi_l|^2}{2m}\nonumber\\ +(m-\mu+\frac{1}{2}m\omega^2r^2)|\Psi_l|^2],
\end{align}
This  describes a system of harmonic oscillators.  The main effect of the Chern Simons term for $\alpha$  is that the vortex carries charge and hence its density couples directly  to the impurity potential, which was accounted for in deriving Eq \eqref{eq:harmonicaction}. Hence the anyon wavefunction should assume the form \[ \Psi_{anyon}\propto exp\left(-\frac{m\omega r^2}{2}\right),\] with a decay length $\sim \frac{1}{\sqrt{m\omega}}$ (a characteristic length describing  the quantum zero-point motion of the vortex).

Note that for the strong impurity potential, near the impurity potential minima, the CDW order $\rho_{mn}$ obtained from linear response and one-loop expansion in Eq \eqref{eq:cdw} does not hold anymore. It is only valid away from the deep potential minima, as shown in the middle panel of Fig \ref{Fig:CDW}(b).

In summary, as one tunes close to the transition and $m\rightarrow 0$, the decay length of the CDW order $1/m$ is much larger than the spatial extent of the anyon wavefunction $\sim 1/\sqrt{m\omega}$. 
Fig \ref{Fig:CDW}(b) shows the profile of composite boson condensate, charge modulation and localized anyon, with two characteristic length scales $1/m, \sim 1/\sqrt{m\omega}$.

\textbf{Doping or weak magnetic field:}
The vortex, identified with the Laughlin quasiparticle, can also be induced by slightly doping the system or turning on an out-of-plane magnetic field. The incompressible state in  density-magnetic field $n-B$ plane traces a line with slope proportional to Hall conductivity (as per the Strada formula). Doping or turning on $B$ away from this incompressible line will create anyon excitations above the ground state, given the robustness of the fractional quantum hall plateau. 

If a vortex $\Phi_v$ that carries $e/3$ is the lowest energy excitation, doping of $\delta$ electrons per unit cell will induce an anyon density of $3\delta$ per unit cell, and associated CDW patches. It is also possible, depending on microscopic details,  that an anyon with $2e/3$, is the lowest energy excitation carrying electric  charge\footnote{It is in principle possible that even an electron with charge-$e$, is the lowest energy charged excitation. However this would be unlikely if the Chern band has close to  ideal quantum geometry which naturally have Laughlin quasiparticles at low energy.}.  Then doping $\delta$ charge per unit cell  induces $3\delta/2$ patches of quasiparticle excitations, and associated CDW halos. When doping $\delta$ electrons induces $3\delta$ or $3\delta/2$ patches of CDW halos,  this indicates fractionalized particles in the system. 

On the dual vortex side, the doping of electrons, equivalently $\Phi$'s, leads to a magnetic field $\nabla\times\alpha(r)=\delta f(r)$ felt by vortex $\Phi_v$'s. The vortex action is supplemented with a term $\alpha_0 \delta f(r)$ in Eq \eqref{eq:vortexaction} from the self Chern-Simons term of $\alpha$, which induces a nonzero chemical potential $\alpha_0$. The $\delta f(r)$ profile hence should be determined self-consistently. A vortex lattice with CDW pattern $\rho_{mn}$ is hence formed, coexisting with FQAH phase. At low anyon or vortex density, we consider a single vortex near the origin in a potential $\alpha_0(r)=\alpha_0^{external}(r)+\alpha_0^{lattice}(r)$ with contributions both from external fields and impurities, as well as other vortex excitations in the vortex lattice. The potential from other vortices mediates a 'Coulomb' interaction among vortices.

As in the discussion for impurities, we expand the potential around its minima as in Eq \eqref{eq:harmonic}, although now, a magnetic field $\delta f(r)$ is also present. The ground state  in this potential is given by $3$ harmonic oscillator modes with a Gaussian form. $\Psi\propto e^{-m\omega_v r^2/2}$, where $\omega_v$ depends the coefficient in a harmonic approximation of the $\alpha_0(r)$ potential, as well as the magnetic field $\delta f(r)$. Ref \onlinecite{balents_competing} showed the charge density modulation to be independent of the pinning potential of the vortex lattice, unlike the case of impurity pinning where $\rho_{mn}\propto V_{mn}$. Essentially this is because vortex excitations already form a vortex lattice and the $V_{mn}$ associated with spatially oscillating potential acts as off-diagonal perturbation within the degenerate $3$ vortex species $\Phi_v$'s space. The resulting $\rho_{mn}$ is given by degenerate perturbation theory with the effective Hamiltonian within the degenerate space depending on $V_{mn}$ and form factors in Eq \eqref{eq:rhomn}. The eigenvectors depend on $V_{mn}$ though its amplitude depends on the  harmonic oscillator wavefunction solved from Eq \eqref{eq:harmonicaction}, not the overall scale of $V_{mn}$.  The profile of the modulation around each vortex core, from Eq \eqref{eq:rhomn}, is thus given by a product of the harmonic oscillator  wavefunctions for $\Phi_v$'s, i.e. is of the Gaussian form $e^{-m\omega_v r^2}$.  The anyon localized around each vortex is again of the size $1/\sqrt{m\omega_v}$.

In the FQAH phase, magnetic field flux will nucleate anyons from the quantum hall response. The case for small magnetic field is similar to doping where CDW order forms with the vortex lattice.

 For impurity pinning the charge density wave nucleated near the vortex core $\rho_{mn}\propto V_{mn}$ is much weaker than those induced by doping or magnetic field. Thus the dominant CDW patches will be those around localized anyons with a much weaker background of CDW order induced by a weak random potential. 
Note that Ref \onlinecite{2023songintertwined} considered an alternative scenario where commensurate CDW coexists with the FQAH and proposed a direct transition to a trivial CDW insulator. There we do not expect anyons to further indpattern uce charge modulations.

\textbf{Proximate CDW with neutral topological order:}

At lattice filling $2/3$, Ref \onlinecite{song2023phase} also proposed another proximate CDW insulator with coexisting neutral topological order.  The resulting q-GL theory reads,
\begin{align} 
\label{1/3fqhtocdwto}
 {\cal L}  =  \sum_{I = 1}^3 |\left(\partial_\mu - i \alpha _{\mu \i} \beta_\mu\right)\Phi_{vI}|^2 +{\mathcal L}_4 +\cdots  \nonumber\\
 - \frac{3}{4\pi} \alpha d\alpha + \frac{1}{4\pi}\beta d\beta+\frac{1}{2\pi} Ad(\alpha+\beta),
\end{align}
where $\alpha, \beta$ are both dynamical $U(1)$ gauge fields. Condensing the $\Phi_v$'s  locks $\alpha=-\beta$ leading to a CDW insulator coexisting with a neutral topological sector described $U(1)_2$ Chern-Simons terms. 

If this scenario is realized, in the FQAH phase, a vortex $\Phi_v$ will carry $2/3$ units of electric charge instead of $1/3$ charge. All the discussion in the manuscript can otherwise be applied to this q-LG theory - essentially the CDW halo around anyons from intertwinement of $2$ phases is predicted by both theories. Unfortunately this also means that this probe can not distinguish the two scenarios. The clear way to achieve this would be a combined measurement of electric and thermal Hall conductivity\cite{song2023phase}.

\section{Summary and experimental probes}
A natural competitor to FQAH ordering at rational fractional filling of a Chern band is a CDW insulator. Indeed numerical calculations often find that these two states are close in energy. In the context of moir\'e systems like twisted MoTe$_2$ or pentalayer rhombohedral graphene, phase transitions between FQAH and a CDW insulator can be driven by small changes in the displacement field. Thus it is only natural that when FQAH order is suppressed in some local region of space, CDW order rears its head. A simple way to suppress the FQAH locally is to introduce an anyon defect. The quantum Ginzburg-Landau theory of Ref. \onlinecite{song2023phase} enables a theoretical treatment of this effect, as we demonstrated in this paper. 

The best experimental probe of the nucleation of density wave order around anyon defects is through Scanning Tunneling Microscopy (STM). However the delicate charge modulation pattern in the FQAH layer and the requirement of high spatial resolution might makes it challenging to observe the effects we predict. In ref \cite{crommie2021}, the charge ordering pattern associated with commensurate Wigner crystals in some moir\'e TMDs was imaged using a novel technique where a graphene sensor layer, separated by a thin hBN,  is placed on top of the TMD bilayers. The  charge modulation of the  of the TMD layer induces associated charge modulation in the graphene layer.   Coupling the tip of an STM to the graphene sensing layer then enables observation of this induced charge modulation. This procedure results in a small perturbation to the TMD system while achieving high spatial resolution. Such a measurement technique is a promising setup to probe the density wave halo around the anyons in FQAH phases.

 Our results  potentially also apply to twisted bilayer graphene/hBN system where a CDW state occurs at zero  magnetic field, and is replaced by a fractional Chern insulator in a small magnetic field\cite{xie2021fractional} at a fractional filling of the moir\'e lattice. Thus clearly there is a close competition between FCI and CDW in this system as well. A small doping of the fractional Chern insulator will induce anyons with CDW halos as an indication of the intertwinement we have discussed. 

The twisted TMD  or twisted bilayer graphene/hBN system will presumably be better suited for this probe than pentalayer graphene, as the FQAH phase occurs at much smaller displacement fields which can presumably also be obtained using the sensor layer as a top gate. Recent theoretical work\cite{dong2023theory,zhou2023fractional,dong2023anomalous} proposes other rhombohedral graphene multilayers as possible platforms for FQAH physics, and, some of these may also offer opportunities to explore the physics discussed in this paper. 

We also note that the extended CDW halo around anyons will complicate coherent  physical braiding processes  that one might want to do to probe their statistics (as the moving the anyons around will require also moving the large region of CDW order). 

Finally should a CDW order be observed at some fraction (for, eg.at $1/3$ filling in twisted MoTe$_2$, or by tuning the displacement field at other fillings), it may be interesting to study the fractionally charged defects of the CDW order. As we discussed, these defects will have an orbital magnetization that reflects the proximity to the FQAH phase. This local change of orbital magnetization compared to the region away from defects may potentially be detected using a scanning nano-SQUID\cite{vasyukov2013scanning} probe.

\emph{Acknowledgements:} We thank Seth Musser and Kevin Nuckolls  for discussions.TS was supported by NSF grant DMR-2206305, and partially through a Simons Investigator Award from the Simons Foundation. XYS was supported by the Gordon and Betty Moore Foundation EPiQS Initiative through Grant No. GBMF8684 at the Massachusetts Institute of Technology. This work was also partly supported by the Simons Collaboration on Ultra-Quantum Matter, which is a grant from the Simons Foundation (Grant No. 651446, T.S.).
%

\end{document}